\documentclass[fleqn,usenatbib]{mnras}

\usepackage{newtxtext,newtxmath}
\usepackage[T1]{fontenc}

\DeclareRobustCommand{\VAN}[3]{#2}
\let\VANthebibliography\thebibliography
\def\thebibliography{\DeclareRobustCommand{\VAN}[3]{##3}\VANthebibliography}

\usepackage{graphicx}	% Including figure files
\usepackage{amsmath}	% Advanced maths commands
\usepackage{booktabs}
\usepackage{color,soul}
\usepackage{tcolorbox}
\title{Pulsar glitches in the presence of vortex traps}

\author[Anantharaman et al.]{S V Anantharaman$^{1}$\thanks{E-mail: ananth.academic@gmail.com},
Dipankar Bhattacharya$^{1}$\thanks{E-mail: dipankar.bhattacharya@ashoka.edu.in}, M. Ali Alpar$^{2}$\thanks{E-mail: ali.alpar@sabanciuniv.edu}, Erbil G\"{u}gercino\u{g}lu$^{3}$\thanks{E-mail: egugercinoglu@gmail.com}
\\
$^{1}$Department of Physics, Ashoka University, Sonipat, 131029, India\\
$^{2}$Faculty of Engineering and Natural Sciences, Sabancı University, Orhanlı, Tuzla, 34956, Istanbul, Turkey\\
$^{3}$ School of Arts and Sciences, Qingdao Binhai University, Huangdao District, 266555, Qingdao, People’s Republic of China\\
}
\date{Accepted XXX. Received YYY; in original form ZZZ}

\pubyear{\the\year{2025}}

\begin{document}
\label{firstpage}
\pagerange{\pageref{firstpage}--\pageref{lastpage}}
\maketitle

% Abstract of the paper
\begin{abstract}
Pulsar glitches are thought to originate when angular momentum is transferred to the crust of the neutron star from the superfluid
enclosed within, mediated by vortex avalanches. This idea has been qualitatively validated in the existing literature by simulating
a star with a small number ($\sim 10^{3}$) of superfluid vortices, subject to deceleration and uniform pinning. Here, we employ the
Barnes-Hut approximation to simulate up to $10^{5}$ vortices in a reasonable time. Using the new setup, we probe glitches that
originate in the presence of inhomogeneous pinning. The inner crust of a neutron star is expected to crack as the star spins
down, relieving stresses and resulting in rearrangements within the crustal lattice. These regions become centres for pinning
with large pinning energies. Many such vortex traps are expected to exist in mature pulsars like Vela. We simulate one such star
and find that the rise of the glitch is now staggered, a clear signature of the trap network. In young to middle-aged stars like PSR
J0537-6910, we expect quakes which result in new traps and also unpin the vortices in existing ones. We observe that such
a configuration involving the simultaneous release of several traps could introduce a bimodality in the glitch-size distribution,
a feature that has previously been reported for PSR J0537-6910. Our simulations also indicate that macroscopic inhomogeneities in the distribution of vortices could spontaneously develop in a star with uniform pinning sites.

\end{abstract}

% Select between one and six entries from the list of approved keywords.
% Don't make up new ones.

%%%%%%%%%%%%%%%%%%%%%%%%%%%%%%%%%%%%%%%%%%%%%%%%%%

%%%%%%%%%%%%%%%%% BODY OF PAPER %%%%%%%%%%%%%%%%%%

\section{Introduction}

The origin of pulsar glitches is widely attributed to the collective unpinning of superfluid vortices in the inner crust of neutron stars (\cite{anderson1975}). At its simplest, the vortex avalanche model posits that an array of quantized vortices and its interaction with pinning structures serve as the basis of most glitches. A complementary explanation has been the crustquake model, whereby the release of crustal stresses powers the observed increase in rotation rate (\cite{ruderman1969, baym1971}). Neither of these ideas, treated independently, has been successful in explaining the host of glitch characteristics observed across pulsars. For instance, in the Crab pulsar, vortex unpinning predicts too short an interval between glitches than what is observed, and quake models predict fractional increases in the rotation and the spin-down rates  that are inconsistent with observations (\cite{alpar1996}). In the case of the Vela pulsar, the initial few glitches fit well into the pure avalanche picture. But prior to the 2016 glitch, the star exhibited a striking decrease in the crustal rotation rate (\cite{palfreyman2018}). This is best explained by a quake in the crust leading to the formation of a trap, a volume having an enhanced vortex density surrounded by a region that is free of vortices (\cite{gugercinoglu2020}).

Additionally, simulations of neutron stars with uniform pinning, where glitches are driven solely by spin-down, mainly suggest a scale invariant power-law glitch-size distribution (\cite{warszawski2008, warszawski2011, howitt2020, liu2024}). However, observations of various pulsars suggest log-normal, Gaussian, and mixed distributions as well (\cite{espinoza2014, howitt2018, fuentes2019}). This diversity can be reconciled only if the star contains structures having a well defined scale. Vortex traps naturally set such a length scale.

It is clear that, in its current form, the vortex avalanche picture neither entirely eliminates contenders such as quake models nor includes them cohesively. The variety in observations and the incompleteness in theoretical understanding motivate us to combine the effects of quakes, traps, and avalanches, yielding behaviour consistent with observed pulsar glitches. A similar approach has been suggested recently by \cite{layek2025}.

Using our setup, we probe various characteristics of glitches in the presence of vortex traps, with a focus on the distribution of glitch sizes. In this study, we have taken most of the neutron star core superfluid to be effectively a part of the normal matter crust to which it is tightly coupled on timescales short compared to glitch rise times, due to the spontaneous magnetization of vortex lines in the core superfluid (\cite{alpar1984}). For the purposes of our simulations, the superfluid outer core of the neutron star where vortex lines will pin and unpin against a toroidal array of flux lines (\cite{gugercinoglu2014}) is not distinguished from the inner crust superfluid where vortices pin to sites in the solid crust.

The paper is organized as follows. In Section 2, we establish a computational setup to simulate a neutron star housing a superfluid with quantized vortices. We also present the relevant performance benchmarks. In Section 3, we discuss vortex traps, their implementation within the simulator, and also introduce quake-like effects within the trap network. Section 4 details the resulting glitch statistics. Section 5 illustrates that significant macroscopic inhomogeneities in the vortex distribution, having a well-defined length scale, could develop spontaneously even in the absence of vortex traps. And in Section 6, we conclude with a discussion and a summary of our work. 

\section{Simulator}

\cite{anantharaman2025} had presented a two-dimensional superfluid vortex simulator based on the initial work by \cite{howitt2020}. Here, we briefly review the relevant setup and highlight the new modifications that allow us to achieve a significant boost in performance. The units for all physical quantities mentioned below are the same as those in the two references stated above.

\subsection{Basic construction}

A neutron star (assumed to be a cylindrical `container' of superfluid) of radius $R$ with a background distribution of pinning sites and almost uniformly distributed vortices sets the stage where the dynamics unfolds. The interaction between the vortex and the bulk superfluid is endowed with a small but finite dissipation, and the star experiences constant deceleration due to the pulsar spin-down torque. The positions of the vortices, then, evolve according  to

\begin{equation}
    \frac{\mathrm{d}}{\mathrm{d} t}\left(\begin{array}{l}
x_{i} \\
y_{i}
\end{array}\right)=\mathcal{R}_{\phi}\left(\begin{array}{c}
v_{i, x} \\
v_{i, y}
\end{array}\right),
\label{eq:eom}
\end{equation}
where
\begin{equation}
    v_{i, x}=-\sum_{j \neq i} \frac{\kappa y_{i j}}{r_{i j}^{2}}+\sum_{j=1}^{N} \frac{\kappa y_{i j, \text { image }}}{r_{i j, \text { image }}^{2}}+\Omega_c y_{i}-\sum_{pin} \frac{\partial V\left(\mathbf{x}_{i}-\mathbf{x}_{pin}\right)}{\partial y_{i}},
    \label{eq:vx}
\end{equation}
and
\begin{equation}
    v_{i, y}=\sum_{j \neq i} \frac{\kappa x_{i j}}{r_{i j}^{2}}-\sum_{j=1}^{N} \frac{\kappa x_{i j, \text { image }}}{r_{i j, \text { image }}^{2}}-\Omega_c x_{i}+\sum_{pin} \frac{\partial V\left(\mathbf{x}_{i}-\mathbf{x}_{pin}\right)}{\partial x_{i}} .
    \label{eq:vy}
\end{equation}
Here, $\mathbf{x}_{i}$ is the position vector of the vortex, whose components are $x_{i}$ and $y_{i}$, calculated with respect to the origin at the centre of the container. $2\pi\kappa$ is the quantum of circulation attached to each vortex, and $\Omega_{c}$ is the angular speed of the container. The terms on the right hand side in equations~(\ref{eq:vx}) and~(\ref{eq:vy}) account for the effect of other vortices, the presence of a boundary imposed via image vortices, the rotation of the star, and the influence of the pinning sites, in that order, on the vortex under consideration. The rotation matrix, $\mathcal{R}_{\phi}$, accounts for the dissipative interactions in the system, which result in a slight radially outward motion of free vortices, described by the small angle $\phi$ with respect to their fast azimuthal motion. The positions of the vortices in the star are evolved by numerically integrating the equations, given the initial configuration of the array. 

The angular speed of the superfluid is related to the distribution of the vortices through
\begin{equation}
\Omega_s=\frac{k}{I_s} \sum_{i=1}^N\left(R^2-r_i^2\right),
\label{eq:superfluidrate}
\end{equation}
where $k$ is a constant fixed by the vortex positions and rotation rate of the superfluid at the start of the simulation. The evolution of the rotation rate of the container is described by
\begin{equation}
\frac{d \Omega_c}{d t}=\frac{N_{\mathrm{ext}}}{I_{c}}-\frac{I_s}{I_c} \frac{d \Omega_s}{dt},
\label{eq:crustrate}
\end{equation}
where $N_{\mathrm{ext}}$ (a negative quantity) is the spin-down torque acting on the star (set to be a constant since it varies only over the pulsar characteristic age, much longer than the simulation's timescale), $I_{s}$ is the moment of inertia of the pinned superfluid, and $I_{c}$ is the moment of inertia of the normal component along with other stellar components coupled to the neutron star crust on rather short timescales. We fix $I_c$ to be unity and define the ratio $I_{\rm{rel}} = I_{s}/I_{c}$ as a free parameter. In all the simulations presented here, we use $I_{\rm{rel}} = 1$. The actual glitch size scales with $I_{\rm{rel}}$, as seen in equation~(\ref{eq:glitchsizeangmomcons}) below. For a description of the equations and notation, refer to \cite{howitt2020}. 

The simulations occur in two phases, the relaxation and the spin-down. In the relaxation phase, the vortices are uniformly distributed throughout the star, to begin with. Then the system is allowed to evolve and stabilize in the absence of spin-down. At the end of this phase, all the vortices are pinned and their distribution is such that the rotation of the superfluid is in equilibrium with the rotation of the star. In the next phase, the spin-down of the star is switched on.

From our simulations, we learn that the system attains equilibrium (relaxes) in a duration that is about one-hundredth the duration of the spin-down run. However, as a general prescription, we set the relaxation run to last one-tenth the duration of the spin-down phase to ensure that all vortices are pinned before the spin-down begins.

\subsection{Barnes-Hut algorithm}

Earlier works had simulated a star with $2\times10^{3}$ vortices and $2\times10^{4}$ pinning sites, with each site having a strength $V = 2\times10^{3}$ (in dimensionless units), just enough for all the vortices to pin during the relaxation phase. The consideration of a trap network demands a crucial improvement to this setup.

A vortex trap in a neutron star is expected to have a cross sectional area of roughly $100~\mathrm{m}^{2}$ (G\"{u}gercino\u{g}lu \& Alpar (2026), in preparation). In a star with a total of $10^{17}$ vortices, a trap could have about $10^{11}$ vortices within it (See Section 3). On the other hand, in the simulations, if we assume 100 traps situated across the model star, each trap would contain a mere 20 vortices, on average. For more traps, the number becomes negligible. Thus we need a system with a larger number of vortices, in order to faithfully approach the real conditions. 

The bottleneck in scaling the prescription in Section 2.1 to a large number of vortices is the $O(N^{2})$ calculation of the vortex-vortex interactions. To overcome this, we suitably adapt the Barnes-Hut (BH) procedure to the present case (\cite{barnes1986}). Similar approaches have been employed to simulate laboratory superfluids, especially in the context of Vortex Filament Models (\cite{hanninen2014}). 

The BH scheme was originally introduced to efficiently simulate several massive bodies interacting with each other gravitationally. The essential insight is that bodies far away from the point of interest contribute only a small force and hence distant clusters can be approximated using an effective mass and centre-of-mass, whereas nearby bodies need to be accounted for exactly.

Since the velocity field set up by the vortices in our system falls with the distance from the vortex as $1/r$, the maximal long-range dependence in 2-D, analogous to the $1/r^2$ dependence of the gravitational force in 3-D, the Barnes-Hut procedure may be applied with suitable modifications. It must be highlighted that treating a neutron star with boundary conditions imposed by image vortices is conceptually a little different from the gravity problem. In fact, it is analogous to the problem of charges interacting via the Coulomb force.  Here, the real vortices and image vortices must be treated as two different species since their circulations have opposite signs. The vector sum of the two provides the net effect on any given vortex. Also, the real vortices are bounded within the star, whereas the image vortices could be present at any arbitrary radius. In practice, a cut-off is applied such that most image vortices are within the cut-off distance.

An outline of the case-specific algorithm is presented below, aimed at capturing the spirit of the procedure rather than the exact details. Our approach can be summarized as follows.

\begin{figure}
  \centering
  \includegraphics[width=\linewidth]{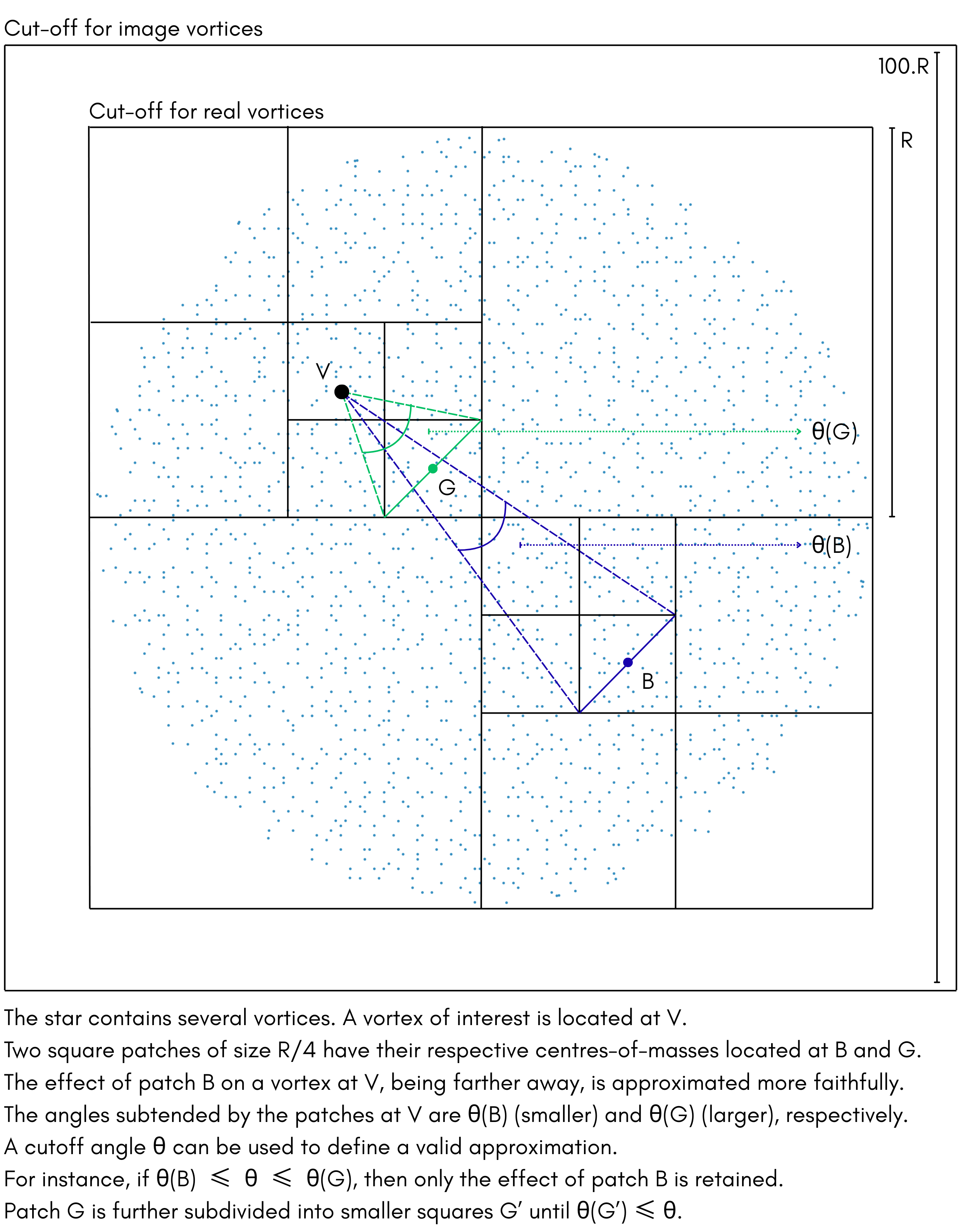}
  \caption{A schematic illustrating the role of the parameter $\theta$ in our adaptation of the Barnes-Hut algorithm. The lower the value of $\theta$, the closer are the results to the exact calculation. The bounding box of size R indicates the region that is partitioned while approximating the effect of real vortices on a given vortex. The image vortices, up to a distance of 100 R are also partitioned, but are not displayed here to reduce clutter.}
  \label{fig:bh_theta}
\end{figure}

\begin{enumerate}
    \item For every time step of integration, the star is partitioned into four quadrants.
    \item The total number of vortices within each quadrant (mass) and the respective centre-of-mass positions are recorded.
    \item Each quadrant is further subdivided into smaller quadrants till the smallest of them contains exactly one vortex. The mass and centre-of-mass are calculated for all sub-quadrants as well.
    \item The net velocity experienced by a given vortex is approximated to be the sum of the velocities due to nearby vortices, calculated exactly, and the velocities due to faraway entire quadrants, approximated using the associated mass and centre-of-mass.
    \item The sense of distance (near or far) is set by a parameter $\theta$, the ratio of the size of the quadrant to the distance of its centre-of-mass from the vortex of interest. See Fig.~\ref{fig:bh_theta}.
    \item The velocity imparted to the given vortex due to all the image vortices is similarly calculated.
    \item A vector addition of the two velocities conveys the total effect that other vortices in the system have on any one vortex, while respecting the appropriate boundary condition.
\end{enumerate}

\subsection{Benchmark}

Implementing the Barnes-Hut procedure results in a time complexity that scales with the number of vortices as $O(N\log(N))$. The performance of the resulting code is benchmarked in Table~\ref{tab:benchmark}, which reflects the execution times for one full run of the simulation.\footnote{The code was tested on Aryabhatta computer system at Ashoka University, courtesy Prof. Suratna Das. The system has 8 cores (AMD Ryzen 7 7700x) supporting two threads each.}

\begin{table*}
\centering
\caption{Benchmark of the N-body superfluid vortex simulator augmented with the Barnes-Hut scheme. The entry in bold is the simulation considered for the error analysis presented in the text.}
\label{tab:benchmark}
\begin{tabular}{@{}llllll@{}}
\toprule
\textbf{Number of vortices} & \textbf{Pinning strength}   & $\mathbf{\theta}$    & \textbf{Error tolerance of integrator}    & \textbf{Execution time} \\ \midrule
2000    & 2000    & 0.8   & $10^{-5}$ &   10 minutes  \\
2000    & 2000    & 0.5   & $10^{-5}$ &   13 minutes  \\
2000    & 2000    & 0.5   & $10^{-12}$ &  30 minutes  \\
10000   & 10000    & 0.8   & $10^{-5}$ &  50 minutes  \\
10000   & 10000    & 0.5   & $10^{-5}$ &  60 minutes   \\
10000   & 10000    & 0.5   & $10^{-12}$ & 150  minutes  \\
\textbf{10000}   & \textbf{20000}    & \textbf{0.5}   & $\mathbf{10^{-12}}$ & \textbf{150  minutes}  \\
$10^{5}$   & $10^{5}$    & 0.5   & $10^{-5}$ & 13 hours  \\
\bottomrule
\end{tabular}
\end{table*}

We have thus managed to simulate up to $10^{5}$ vortices moving in an environment of a suitably high pinning strength, in significantly less time compared to the figure of more than one month reported in \cite{cheunchitra2024} for a simulation with $5\times10^{3}$ vortices. The validity of these simulations is checked against exact calculations by comparing the vortex array obtained at the end of the relaxation phase in both cases. \cite{howitt2020} provide a measure to capture the state of the array by associating an energy defined as

\begin{equation}
    \mathcal{H}=\sum_{i=1}^N \sum_{j \neq i} \kappa \ln \left(r_{i j}\right),
    \label{eq:arrarenergy}
\end{equation}

where $r_{ij}$ is the pair-wise distance between vortices and $\kappa$ is the quantum of circulation. Using this, we define a percentage error in energy to compare the final states of systems that started with the same initial set of parameters, differing only in the force calculation scheme:

\begin{equation}
\Delta\mathcal{H} = \frac{\mathcal{H}_{\text{Barnes-Hut}} - \mathcal{H}_{\text{Exact}}}{\mathcal{H}_{\text{Exact}}} \times 100.
\label{eq:percentageerror}
\end{equation}

For the simulation with 10000 vortices and pinning strength $V = 20000$, the percentage error $\Delta\mathcal{H}$ is $6\times10^{-3}$, lending credibility to the use of the Barnes-Hut procedure in simulating point vortices. Further, we also inspect the final positions of the vortices visually and find that only a small fraction are out of agreement, leading to negligible effects on the statistics of the system.

Our code is an augmented reimplementation, written in C++, fully independent of the one presented in \cite{howitt2020}. It is also modular, enabling easy addition and activation of various functionalities. The Data Availability section of this article contains a link to the pertinent code. The simulations in the following sections are run with a modest initial population of $5 \times 10^{3}$ vortices, $\theta = 0.5$ and an error tolerance of $10^{-5}$, striking a positive balance between performance and numerical accuracy.

\section{Vortex traps in neutron stars}

The inner crust of neutron stars is expected to break due to stresses associated with spin-down and vortex pinning, leading to ``vortex traps'' with large pinning energies (\cite{alpar1996}). Such traps have an excess density of vortices, creating a strong circulation around the trap. This circulation, alongside the comparatively weak pinning outside the trap, creates a vortex-free region (\cite{cheng1988}). The crust of neutron stars can then be imagined as a network of traps, each of which is necessarily surrounded by a vortex free zone.

The size of the traps is determined by the crystal properties. The critical strain angle for the inner crust is expected to be between 0.1 and 0.01 (\cite{horowitz2009, baiko2018, caplan2026}). If the traps extend across the thickness of the crust ($\sim 1~\mathrm{km}$), then the strain angle implies a maximum cross sectional area between $10^{2}$ and $10^{4}~\mathrm{m}^{2}$. A more involved calculation is presented in \cite{gugercinoglu2019}, where an estimate closer to $10^2~\mathrm{m}{^2}$ is obtained. They also suggest that approximately $10^{3}$ such regions should be involved to explain the largest glitch of the Crab pulsar.

To capture the essence of such a picture, we construct a star with a strong separation of high vortex density regions (traps) and vortex-free regions (voids). This is a simple model that captures the idea described above. The traps have strong pinning sites and the voids have no pinning sites. The size of a trap and the distance between the centre of neighbouring traps are both parameters that can be set by the user. Here, as a simple case, we take the distance between the traps to be twice the size of the traps. The various parameters relevant to all the simulations presented in the following sections are summarized in Appendix A, with detailed comments on the chosen values.

To begin with, we initialize all the vortices within the traps. The initial rotation rate of the star is set to a value close to what a uniform array containing the same number of vortices would have. During the relaxation phase, several vortices are ejected out of the traps and into the vortex-free regions. By the end of this phase, many of them exit the system. The resulting equilibrium configuration is such that all the remaining vortices are pinned within the traps and experience zero net force. After this, we begin the spin-down which leads to an increase in the lag between the superfluid rotation rate, which is kept constant by pinning, and the crust-normal matter rotation rate which is decreasing. The lag is sustained by the available pinning forces. When the `Magnus force' resulting from the growing lag exceeds the maximum available pinning forces, the simulation results in several glitches.

\begin{figure}
  \centering
  \fbox{\includegraphics[width=0.97\linewidth]{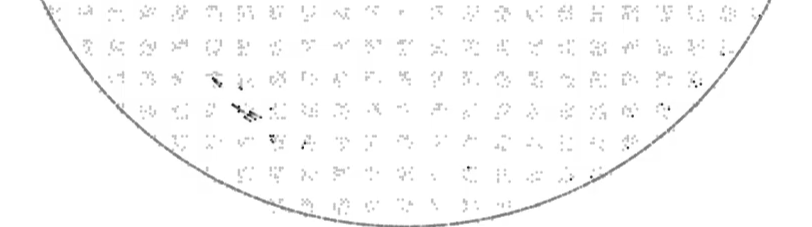}}
  \fbox{\includegraphics[width=0.97\linewidth]{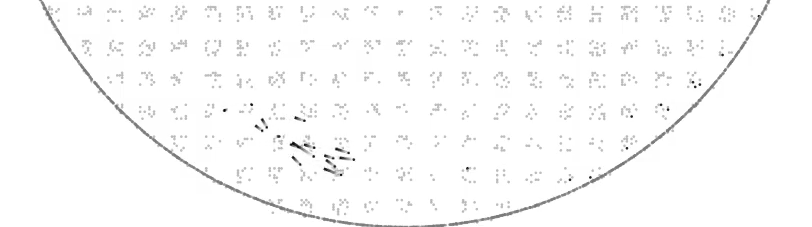}}
  \fbox{\includegraphics[width=0.97\linewidth]{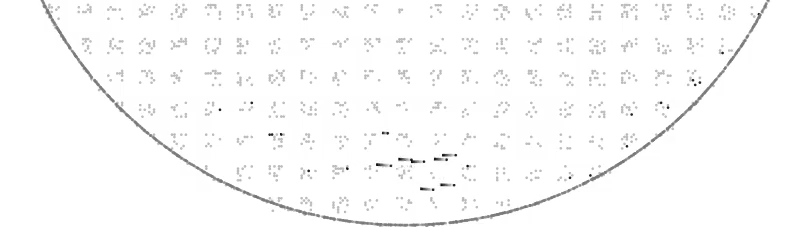}}
  \fbox{\includegraphics[width=0.97\linewidth]{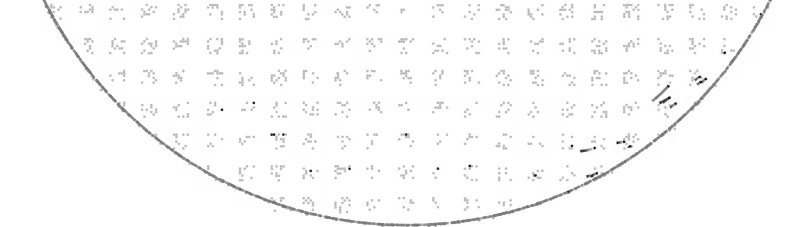}}
  \caption{Four chronologically arranged snapshots of a glitch in the presence of traps, occurring in the bottom quarter of a star. The glitch lasts for a total of $0.8~T_{0}$. The four panels show the vortex array at $0.1~T_{0}$, $0.2~T_{0}$, $0.4~T_{0}$, and $0.7~T_{0}$ from the start of the glitch. Panel 1: Some vortices from a trap are released due to the Magnus force exceeding the pinning force. Panel 2: Vortices from neighbouring traps are released due to the displacement of those in the first trap. Panel 3: Some of the previously unpinned vortices have repinned. The rest continue to move through the vortex-free spaces. Panel 4: Release of vortices from a few traps, closer to the boundary of the star, ensues. Some vortices annihilate on reaching the boundary of the superfluid region.}
  \label{fig:schematic_traps}
\end{figure}

\subsection{Characteristics of the resulting glitches}

Four illustrative snapshots of a glitch, chronologically arranged and occurring in the bottom quarter of a star with about 500 traps, are presented in Fig.~\ref{fig:schematic_traps}. The black dots are all the vortices that participate in the glitch. The grey dots represent the vortices that remain pinned. The tails attached to the black dots represent movement, and their lengths convey the instantaneous speed of the vortices. The pinning sites are not marked, but the waffle-like arrangement of the vortices points to the location of the traps.

During the initial constant spin-down, we find that vortices rearrange themselves within the traps. When the rotational lag between the superfluid and the crust becomes large, and the appropriate local conditions are met, a few vortices exit from a trap and travel across the vortex-free region. They stimulate the release of more vortices from other traps and the process continues till the lag decreases. Such a behaviour was predicted by \cite{cheng1988} and \cite{alpar1996}.

%Fig 1 previous location%

Further, we notice that the rise of some glitches, especially small ones, is now staggered. This feature encodes the distribution of the trap network. A quantitative probe of this effect shall be explored in a future work. But a qualitative understanding can be easily gained from animation Anim1 (Supplementary Data) which shows the evolution of the vortex array during a typical glitch. It also displays the corresponding change in the rotation rate of the container. Note that all the glitches in our simulations rise fast (spanning roughly one rotation period) compared to the spin-down (lasting about two thousand rotation periods).

\subsection{Triggered release}

There is a lack of consensus about the exact origin of a glitch. The diverse observational features exhibited by glitches in the same pulsar have led to the suggestion that several mechanisms could be acting alongside each other (\cite{antonopoulou2022, zhou2022}). The earliest explanation for glitches proposed that stress-related crustal failure was the sole reason for the star’s spin-up (\cite{baym1969, ruderman1969}). Although these models did not explain the glitch data corresponding to several pulsars, interest in crust-breaking has resurfaced due to its effect on the vortex array (\cite{haskell2015, rencoret2021}). Such a paradigm may be studied by allowing for perturbations of the pinning sites (triggers) to serve as the impetus for vortex unpinning, as opposed to `spontaneous' unpinning due to superfluid dynamics, when, as a result of spin-down, the Magnus force exceeds the pinning force. A recent work by \cite{layek2025} suggests that we may have to invoke such triggers and consider their effect on the vortex array to be able to explain large glitches that involve up to $10^{13}$ vortices.

\begin{figure}
  \centering
  \fbox{\includegraphics[width=0.97\linewidth]{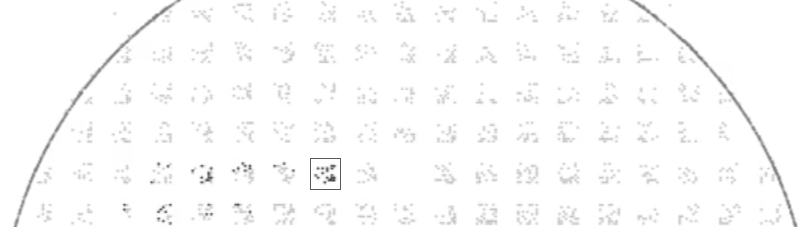}}
  \fbox{\includegraphics[width=0.97\linewidth]{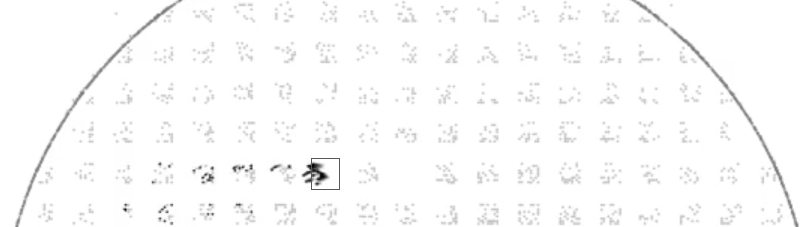}}
  \fbox{\includegraphics[width=0.97\linewidth]{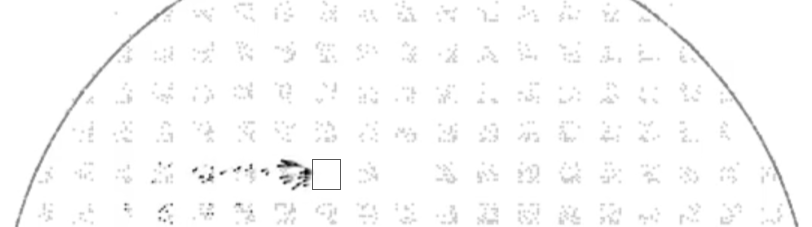}}
  \fbox{\includegraphics[width=0.97\linewidth]{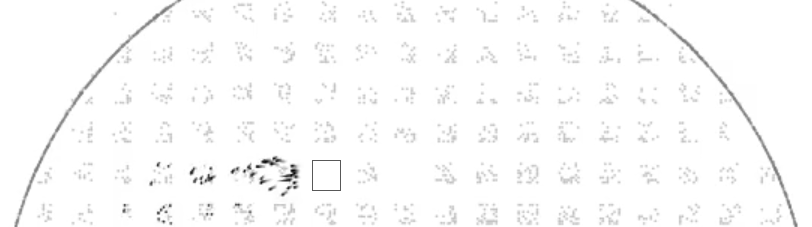}}
  \caption{Four chronologically arranged snapshots of a vortex trap being triggered, spanning a total of $0.15~T_{0}$. Only the relevant section of the star is presented. The four panels show the vortex array at $0.0~T_{0}$, $0.05~T_{0}$, $0.1~T_{0}$, and $0.15~T_{0}$ from the start of the trigger. Panel 1: A randomly chosen trap (marked by a black box) is triggered. Panel 2: A few vortices exit the trap and enter a neighbouring trap. Some get repinned and some lead to further release of vortices. Panel 3: The vortices from the originally triggered trap have all exited the trap. Panel 4:  The evolution of the vortex array continues and a significant release is noticed in the neighbouring trap.}
  \label{fig:schematic_traps_trigger}
\end{figure}

To facilitate such a study, we mimic the effect of a crust rearrangement on a trap network. We switch off the pinning sites within a trap to release all the vortices within it simultaneously. We let the trap become active after a period specified as a free parameter. This duration is usually a small number of the order of a rotation period ($\sim 1~T_{0}$). Four chronologically arranged snapshots of triggered release are shown in Fig.~\ref{fig:schematic_traps_trigger}. All the vortices within a single randomly chosen trap are seen to escape simultaneously, validating the correctness of the computational module.

We implement a mechanism by which a set of traps across the star are turned off a prescribed number of times throughout the run. The choice of the traps that are switched off, and the times at which such a procedure is carried out are all random.

Animation Anim2 (Supplementary Data) highlights the evolution of the vortex array during one such triggered glitch, where a single triggered trap leads to the release of vortices from a chain of traps.

\section{Glitch statistics in the presence of traps}

We run several simulations, as listed below, to understand the distribution of glitch sizes in the presence of vortex traps. The total number of traps present within the star, approximately 300, remains the same in all the cases. The statistics produced in each case are based on the data accumulated from five iterations of the same setup, differing only in the initial configuration of the vortex array.
\begin{enumerate}
	\item One simulation with no triggers imposed on the trap network. Glitches result solely due to spin-down. That is, all the glitches here are spontaneous.
	\item Three simulations, each having 10 triggers at random times during the simulation. Each trigger involves 1 trap, 5 traps, and 10 traps, in the three cases, respectively.
	\item Three simulations, each having 20 triggers at random times during the simulation. Each trigger involves 1 trap, 5 traps, and 10 traps, in the three cases, respectively.
	\item Three simulations, each having 30 triggers at random times during the simulation. Each trigger involves 1 trap, 5 traps, and 10 traps, in the three cases, respectively.
\end{enumerate}

The radial motion of a vortex, even if small, produces a variation in the total angular momentum of the superfluid. This in turn results in a change in the rotation rate of the normal part of the star. Our simulator has an inbuilt glitch-detector which records all such variations in the rotation rate, no matter how small. The rate of change of $\Omega_{c}$ is negative during constant spin-down. A glitch is said to begin when it becomes positive and end when it returns to being negative. The size of the glitch is then defined as the increase in rotation rate in this interval, $\frac{\Delta\Omega_c}{\Omega_0}$. However, only glitches produced by a collective motion of vortices, resulting in a large change in the rotation rate, are observationally relevant. This range of glitch sizes is found empirically from the simulations and a cut-off glitch size is chosen appropriately such that only observationally relevant variations, involving significant collective movement, are accounted for. By an argument similar to the one made in \cite{anantharaman2025}, we set the cut-off to be $3 \times 10^{-3}$. It is this limit that we shall probe in the remainder of this section.

\begin{figure}
  \centering
  \includegraphics[width=\linewidth]{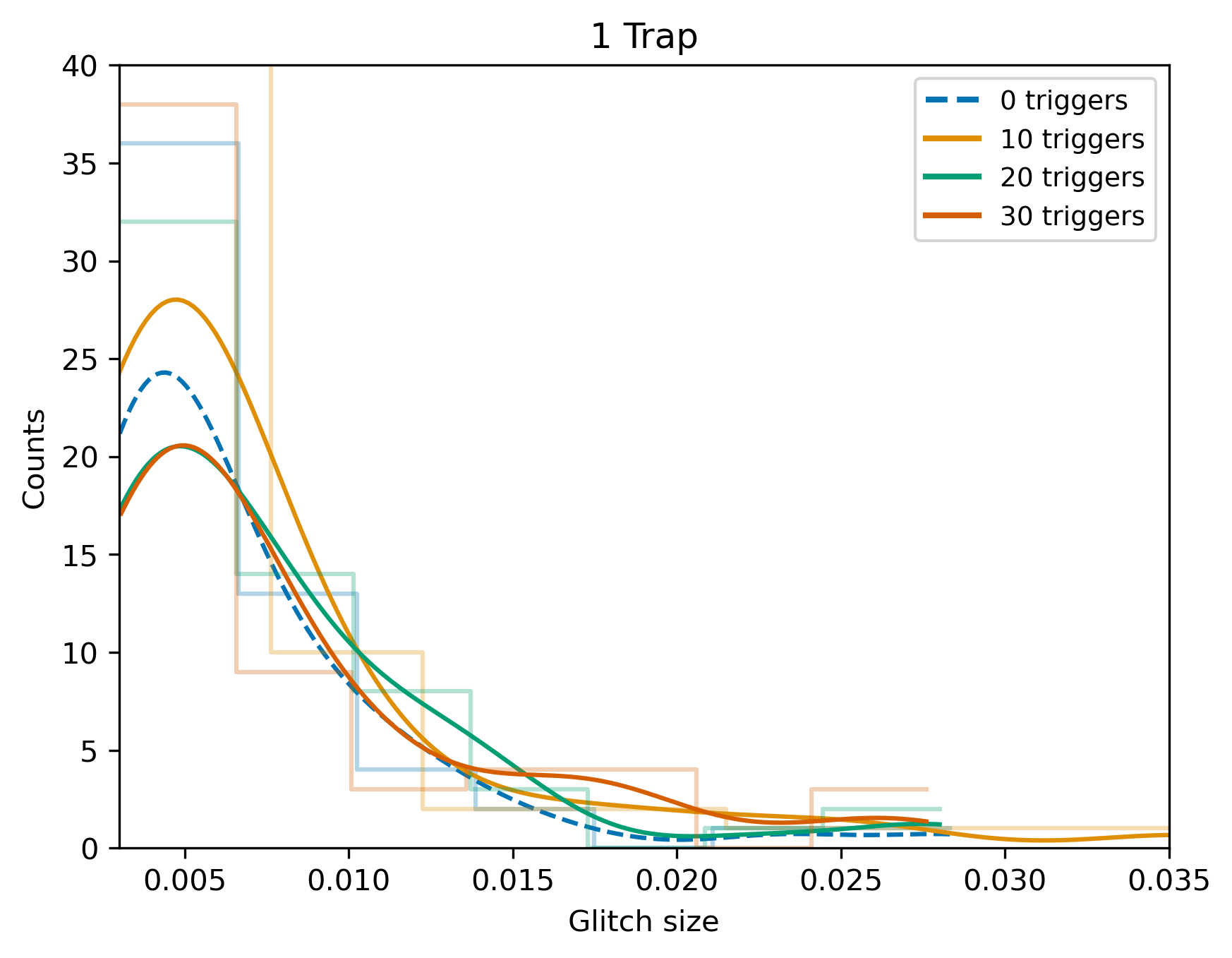}
  \includegraphics[width=\linewidth]{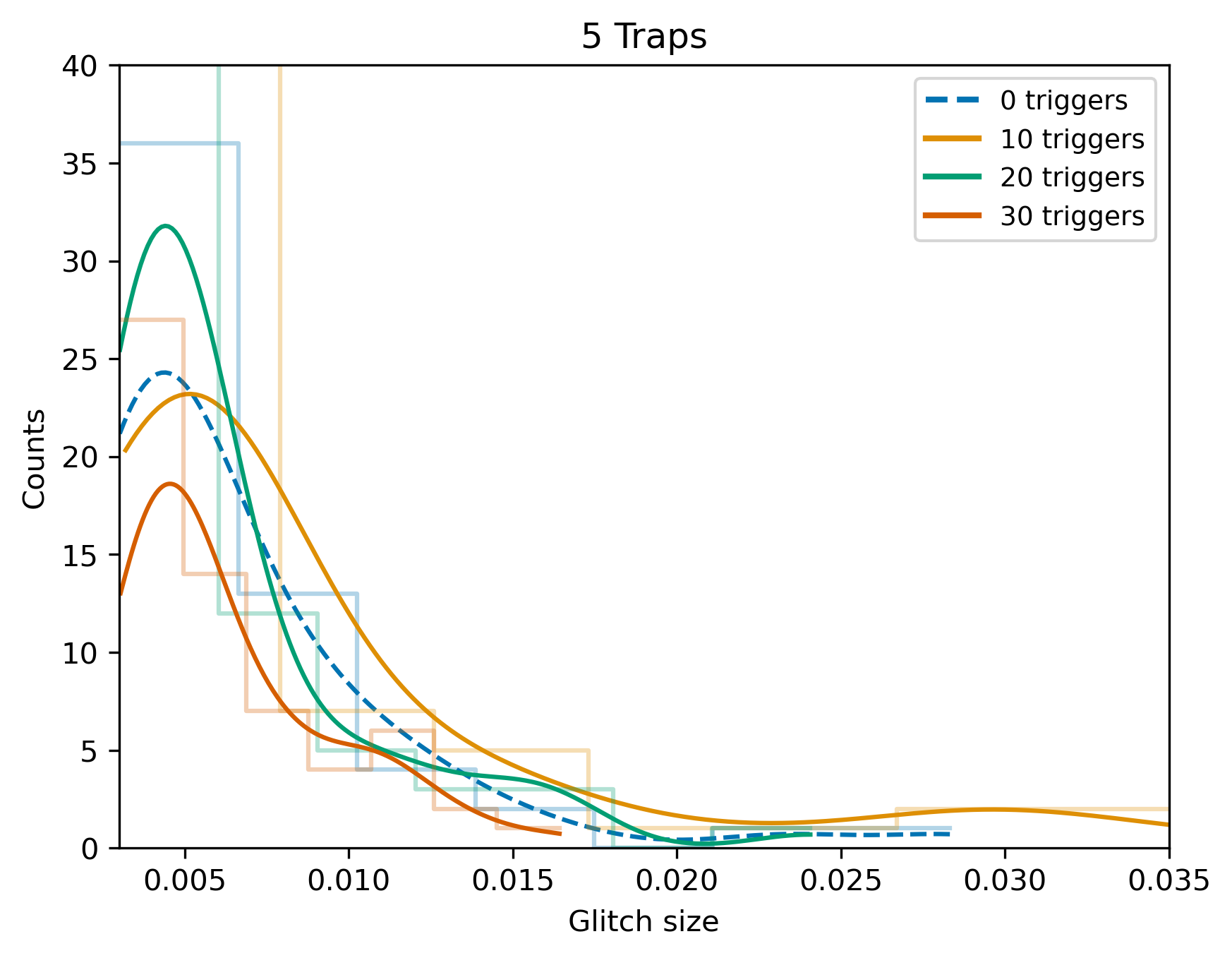}
  \includegraphics[width=\linewidth]{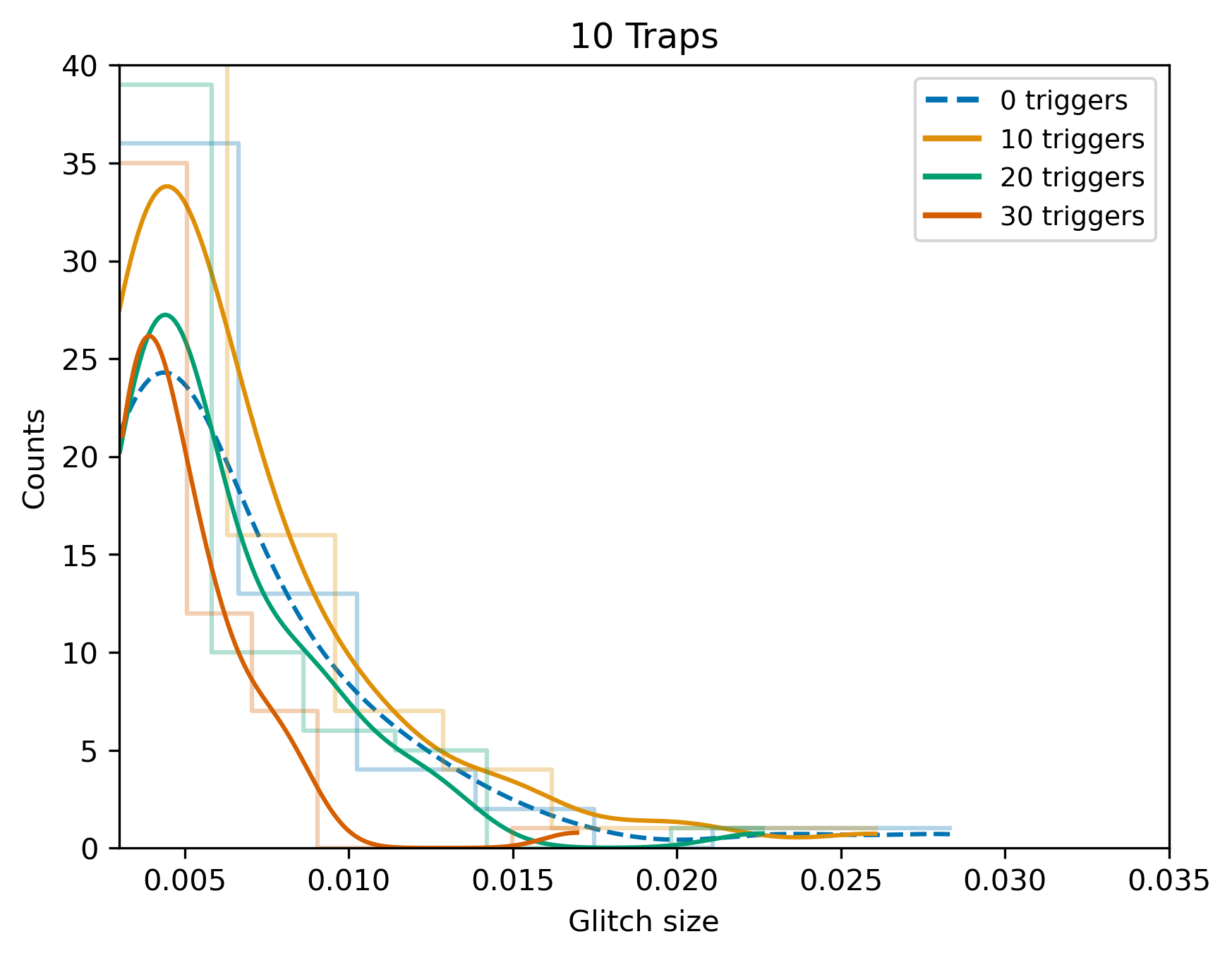}
  \caption{Distribution of glitch sizes resulting from various simulations involving vortex traps. Perspective 1: Each panel considers simulations having the same number of triggered traps, varying only in the number of triggers implemented during the runtime.}
  \label{fig:trapwise_consttraps}
\end{figure}

\begin{figure}
  \centering
  \includegraphics[width=\linewidth]{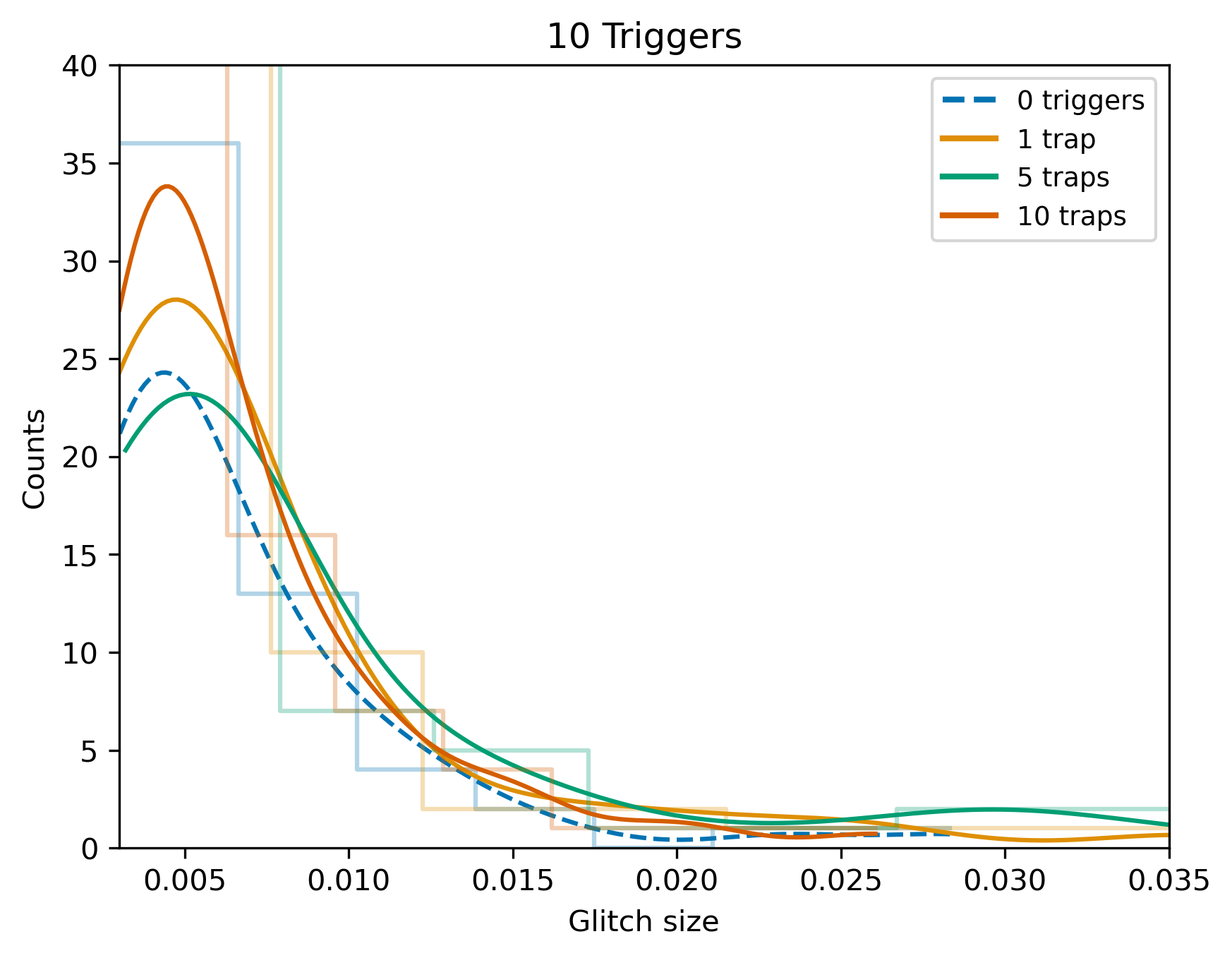}
  \includegraphics[width=\linewidth]{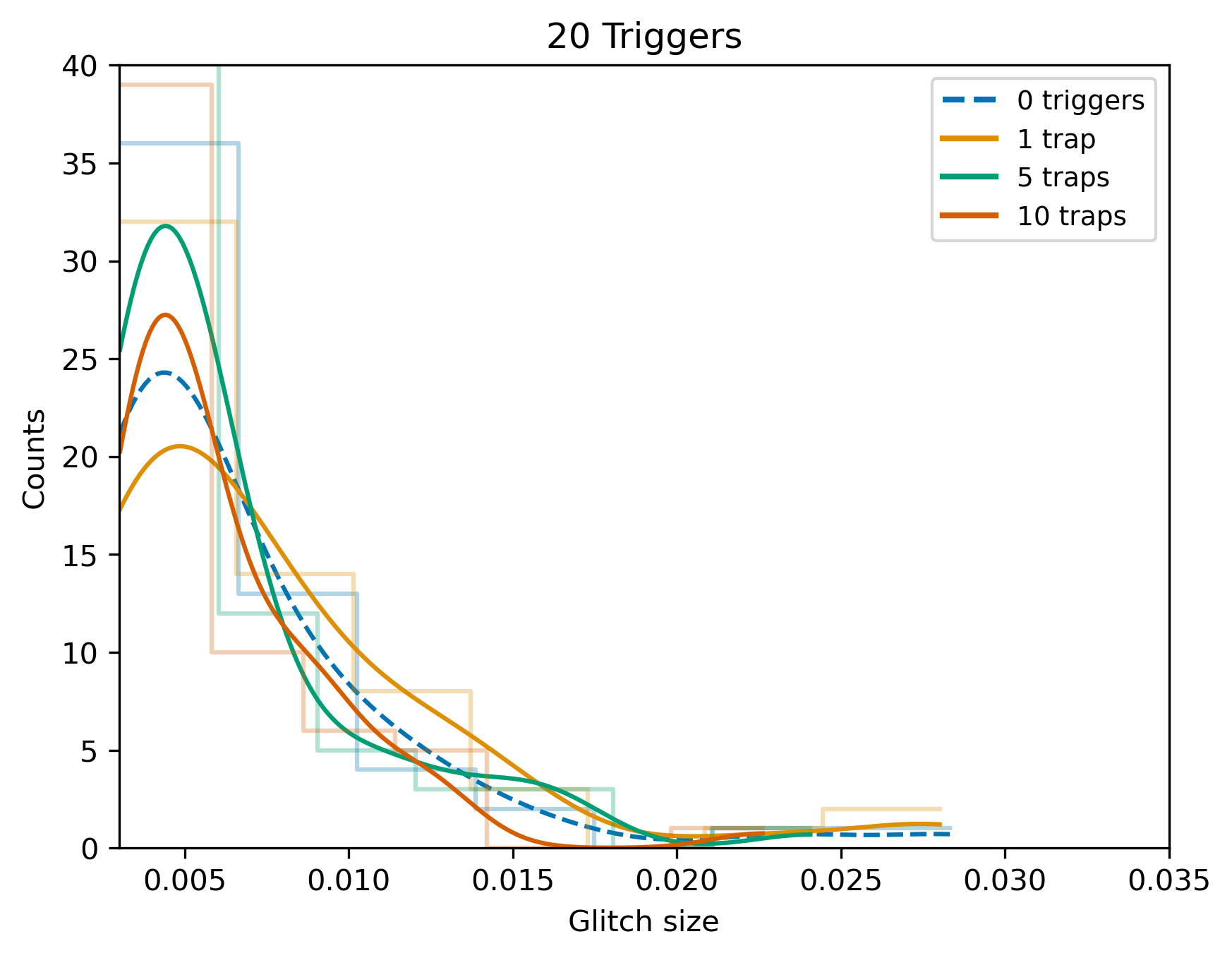}
  \includegraphics[width=\linewidth]{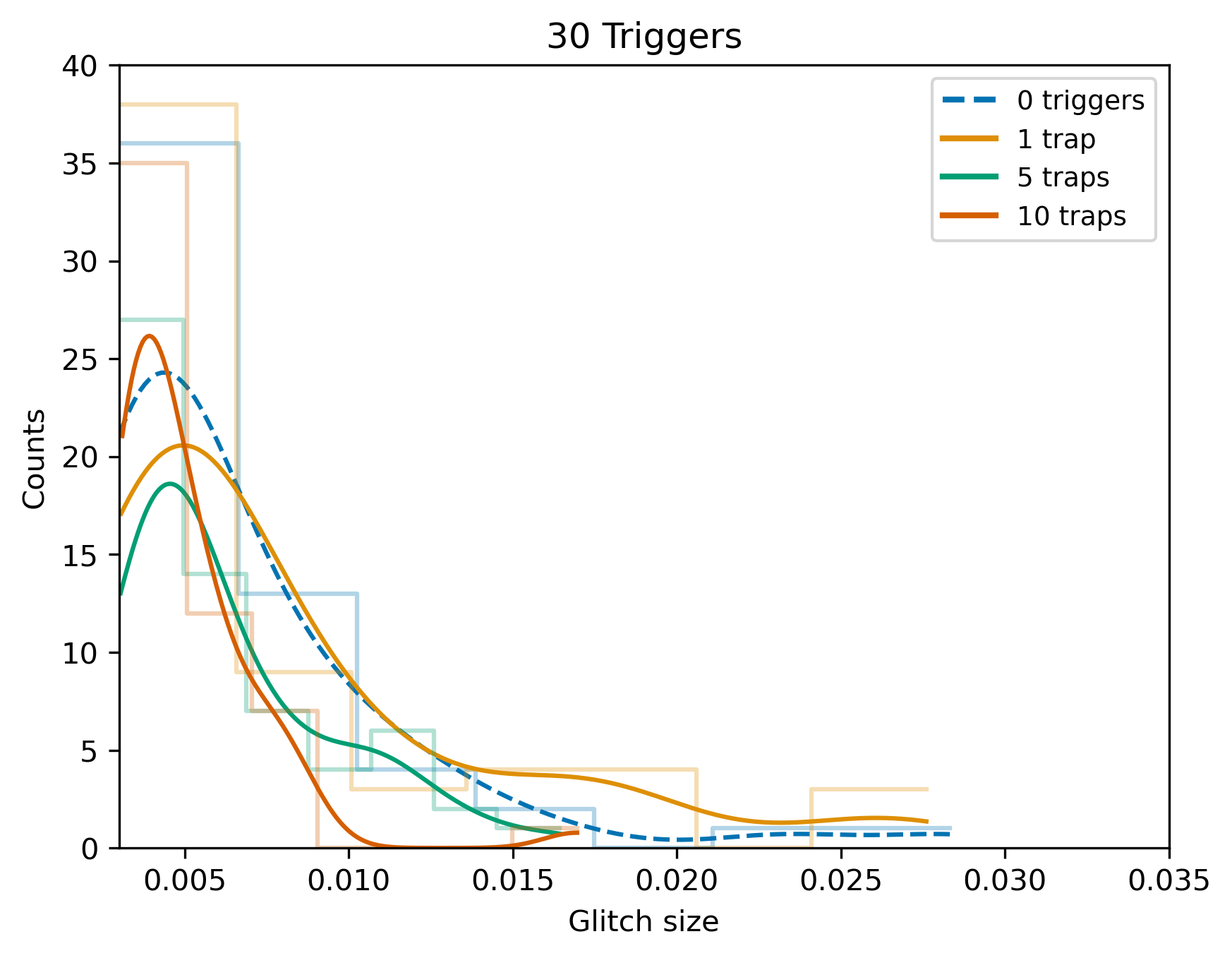}
  \caption{Distribution of glitch sizes resulting from various simulations involving vortex traps. Perspective 2: Each panel considers simulations having the same number of runtime triggers, varying only in the involvement of traps.}
  \label{fig:trapwise_consttriggers}
\end{figure}

The distributions resulting from the above simulations are compared in Fig.~\ref{fig:trapwise_consttraps} and Fig.~\ref{fig:trapwise_consttriggers}. The thick solid (or dotted) lines indicate the corresponding Kernel Density Estimates (\cite{parzen1962}). These figures provide two different perspectives to understand the same set of data. In Fig.~\ref{fig:trapwise_consttraps}, each panel presents simulations having the same number of triggered traps, varying only in the number of triggers implemented during the runtime. Fig.~\ref{fig:trapwise_consttriggers} shows simulations having the same number of runtime triggers in each panel, with different options for the number of traps involved.

\subsection{Analysis}

The panels in Fig.~\ref{fig:trapwise_consttraps} and Fig.~\ref{fig:trapwise_consttriggers}, taken together, suggest that both the number of triggered traps and the number of such triggers are vital in deciding the size of the largest glitch and also the form of the distribution. Prominent secondary peaks start appearing in the tail of the distribution as we increase trap involvement. This behaviour is culled when the number of triggers is large. That is, if the stress built up in the pinned vortex array is released frequently, by invoking many triggers or involving a large number of traps, or both, then the size of the resulting glitches reduces significantly. To understand this, we begin by estimating the size of a glitch that results due to a single trap being triggered.

The superfluid rotation rate associated with any configuration is given by equation~\ref{eq:superfluidrate}, reiterated below:
\begin{equation*}
\Omega_s=\frac{k}{I_s} \sum_{i=1}^N\left(R^2-r_i^2\right),
\label{nothing}
\end{equation*}
where $k$ is a constant fixed by the vortex positions and the rotation rate at the start of the spin-down phase. Assuming that the spin-down due to external torque is negligible during a glitch, conservation of total angular momentum gives the size of a glitch as
\begin{equation}
\frac{\Delta\Omega_{c}}{\Omega_{0}} = -I_{\rm{rel}}\frac{\Delta\Omega_{s}}{\Omega_{0}},
\label{eq:glitchsizeangmomcons}
\end{equation}
where $\Delta\Omega_{s}=\Omega_{s,\rm{end}}-\Omega_{s,\rm{start}}$ is the difference in the rate of superfluid rotation calculated between the end and the beginning of the event.

Let the average number of vortices in a trap be denoted by $n$, the initial location of a vortex be $r_{A}$, and the final location after the trigger and resettlement be $r_{B}$. Since the triggered traps are selected randomly during runtime, on average we can assume the trap to be located at $R/\sqrt{2}$, the radius at which the number of traps within equals the number of traps outside. Further, we consider the location of all the vortices within the trap to be approximately the same before the trigger, thus $r_A = R/\sqrt{2}$. After the trigger, some vortices will get repinned closer to the site of trigger and some will exit the system. Crudely, the vortices can be thought to migrate to a site radially halfway between the origin of the trigger and the boundary of the star. That is, the location at the end of the glitch could be approximated as $r_B = 0.85~R$. In our simulations, we find that the constant $k/I_{s} = 1.6\times10^{-4}$, and the average number of vortices in a trap $n=15$. We set the initial rate of rotation to be $\Omega_{0} = 40$, informed by the case of a superfluid having $N_v$ vortices arranged uniformly\footnote{Note that the quantum circulation in our simulations is given by $2\pi\kappa$, where $\kappa$ is set to unity.}, $\Omega_{\text{u}}= N_v(2\pi\kappa)/2\pi R^{2} = 50$, and tuned to minimize the exit of vortices during the relaxation phase. The glitch size associated with the trigger of a single trap can then be obtained as:

\begin{equation}
\frac{\Delta\Omega_{c}}{\Omega_{0}} = I_{\rm{rel}}~\frac{k}{I_s}~n~\frac{1}{\Omega_{0}}~(r_{B}^{2} - r_{A}^{2}) = 1.4\times10^{-3}.
\label{eq:glitchestimate1trap}
\end{equation}
Glitches involving 5 and 10 traps then possibly result in glitch sizes populated around $7\times10^{-3}$ and $1.4\times10^{-2}$, respectively.

Prior to the simulations, we make the estimates presented above. After the simulations, we compare the resulting distribution with the above predictions. As mentioned earlier, we consider only large glitches resulting from the collective motion of vortices such that $\frac{\Delta\Omega_{c}}{\Omega_{0}} > 3\times10^{-3}$. In the top panel of Fig.~\ref{fig:trapwise_consttraps}, we see that triggering only 1 trap several times during the run alters the total number of glitches recorded. This would not be possible if the triggers produced glitches strictly having a size of $1.4 \times 10^{-3}$ as per our estimate, since this is smaller than the imposed cut-off. The middle panel, concerning triggers involving 5 traps, suggests a low albeit clear second peak around $3\times10^{-2}$ when the array is triggered 10 times during the simulation, again differing from the expectation. Note that such a bimodality, which has previously been reported for PSR J0537-6910 (\cite{celora2020, anantharaman2025}), spans only a restricted range of glitch sizes. This is different from the preference for larger and smaller glitches exhibited by the Vela and Crab pulsars, respectively, where the distribution spans several orders of magnitude. The bottom panel, pertaining to triggers involving 10 traps, as well as the middle and the bottom panels of Fig.~\ref{fig:trapwise_consttriggers}, indicates that the size of the biggest glitch decreases in comparison to those resulting from simulations involving a single trap.

The disagreement between the predictions and the appearance of the second peak in the recorded glitch sizes reflects a gap in the elementary analysis presented above. It fails to capture the fact that triggering a trap could lead to avalanches containing a number of vortices much larger than what was initially housed within it. In our simulations, we notice that a single-trap trigger (releasing 15 vortices) leads to the movement of up to 300 vortices during a glitch. A faithful representation should, then, capture the dependence of avalanches on (i) the number of vortices initially displaced, given by the number of traps triggered ($n_{\rm{traps}}$), and (ii) the average stress stored in the system, which is related to the number of triggers implemented during the runtime ($n_{\rm{triggers}}$). To this end, we modify equation~(\ref{eq:glitchestimate1trap}) by explicitly introducing the number of traps triggered, and a suitable amplification factor ($\gamma$) to account for the average stress. This leads us to
\begin{equation}
\frac{\Delta\Omega_{c}}{\Omega_{0}} = \gamma ~\left[ n_{\rm{traps}}~ I_{\rm{rel}}~\frac{k}{I_s}~n~\frac{1}{\Omega_{0}}~(r_{B}^{2} - r_{A}^{2}) \right].
\label{eq:glitchestimate1trap_new}
\end{equation}
In order to understand how the amplification factor varies across simulations, we consider the size of the largest triggered-glitch in each case as the relevant glitch size in equation~(\ref{eq:glitchestimate1trap_new}). The corresponding values of the amplification factor are collected in Table~\ref{tab:gamma}. Note that $\gamma$ can also be interpreted as a measure of the closeness of the simple estimate provided by equation~(\ref{eq:glitchestimate1trap}) to the actual glitch size resulting from the simulation.

Table~\ref{tab:gamma} immediately suggests that, for a fixed number of triggers, the amplification decreases sharply with increasing trap involvement. Triggers in which only a single trap participates have a high amplification, akin to spin-down induced spontaneous glitches. In scenarios with more triggered traps, the amplification is significantly less, resulting in maximum glitch sizes not too different from the single trap case. Addressing this from first principles shall be pursued in a future work.

\begin{table}
\centering
\caption{Values of the amplification factor $\gamma$ derived from the various simulations presented in this study.}

\label{tab:gamma}
\begin{tabular}{@{}lllll@{}}
\toprule
$\mathbf{n_{triggers}}$ & $\mathbf{n_{traps}}$   & \textbf{Largest triggered glitch $(10^{-3})$}    & $\mathbf{\gamma}$ \\ \midrule
10    & 1    & 23   & 16.4 \\
10    & 5    & 28   & 4.00 \\
10    & 10    & 25   & 1.80 \\
\midrule
20   & 1    & 33   & 23.6 \\
20   & 5    & 23  & 3.30 \\
20   & 10    & 30   & 2.10 \\
\midrule
30   & 1    & 27   & 19.3 \\
30   & 5    & 16   & 2.30 \\
30   & 10    & 17   & 1.30 \\
\bottomrule
\end{tabular}
\end{table}

\subsection{Remarks}

\subsubsection{Dependence on the relative moment of inertia}

The range of glitch sizes in the results presented above is sensitive to the choice of $I_{\rm{rel}}$, which we had fixed to be unity. To check whether this dependence is indeed as informed by equation~(\ref{eq:glitchestimate1trap_new}), we simulate a set of stars with different values of $I_{\rm{rel}}$, keeping $I_c$ fixed at unity. We find that simulations of stars having relative moment of inertia considerably smaller than unity ($I_{\rm{rel}}\sim0.01$), produce significantly fewer glitches for the same runtime, requiring numerous iterations to establish reliable statistics. However, the range of glitch sizes thus obtained for each scenario described above is found to be in line with that suggested by (\ref{eq:glitchestimate1trap_new}).

\subsubsection{Dependence on the trap ratio}

In all of the above simulations, the ratio of the distance between the traps to the size of an individual trap is set to be 0.5. This serves well to highlight the separation, giving the vortex distribution its characteristic waffle-like pattern. However, it is far from the actual value expected in a neutron star. To see this, consider a star with vortices distributed uniformly and pinned completely, such that the superfluid rotation is in equilibrium with the crustal rotation, having an angular speed $\Omega$. Let the critical differential rotation between the two components required for unpinning be $\Omega_{\rm{cr}}$. In the equilibrium configuration, the differential rotation is zero. However, in the presence of traps, the vortices that were previously in an area $D^2$ are rearranged into a smaller area $D'^2$, leading to an increase in the density of vortices within. The local velocity field increases inversely proportional to the area of the trap. Then, the smallest possible trap is one where the vortices remain just pinned. Let the distance between the traps be $D$ and the size of the traps be $D'$. Then the trap ratio $ (D'/D) = \left( 1-(\Omega_{\rm{cr}} / \Omega) \right)^{1/2}$. For a typical pulsar with a rotation rate of $\sim40~\rm{rad~s^{-1}}$, unpinning is expected to occur for a differential of $\sim1~\rm{rad~s^{-1}}$, suggesting a trap ratio of 0.9875. The gap between two traps, each ten meters in size, is then expected to be of the order of ten centimeters. In our simulations, such a configuration will be practically indistinguishable from a uniform distribution. Increasing our ratio to 0.7, keeping the trap size constant, leads to minor changes in the distributions presented. Increasing it further takes the distribution closer to the one corresponding to the uniform case. However, if we also change the size of the traps, keeping the ratio at 0.5, such that the star contains about 500 traps in total, the staggered imprint on the glitch-rise is found to be stronger.

\section{Spontaneous inhomogeneities in a uniform array}

Previous point-vortex studies of the neutron star superfluid featured uniformly distributed vortices against a background of pinning sites. The critical unpinning lag supported by these sites was fixed to be around 7 percent of the star's initial rotation rate. Smaller values of the lag did not produce detectable glitches and larger values were computationally expensive (\cite{howitt2020, cheunchitra2024, anantharaman2025}). The enhanced setup presented in Section 2 allows us to probe a neutron star with not only a large number of vortices, but also with pinning sites supporting a wider range of critical lags.

\begin{figure}
  \centering
  \includegraphics[width=\linewidth]{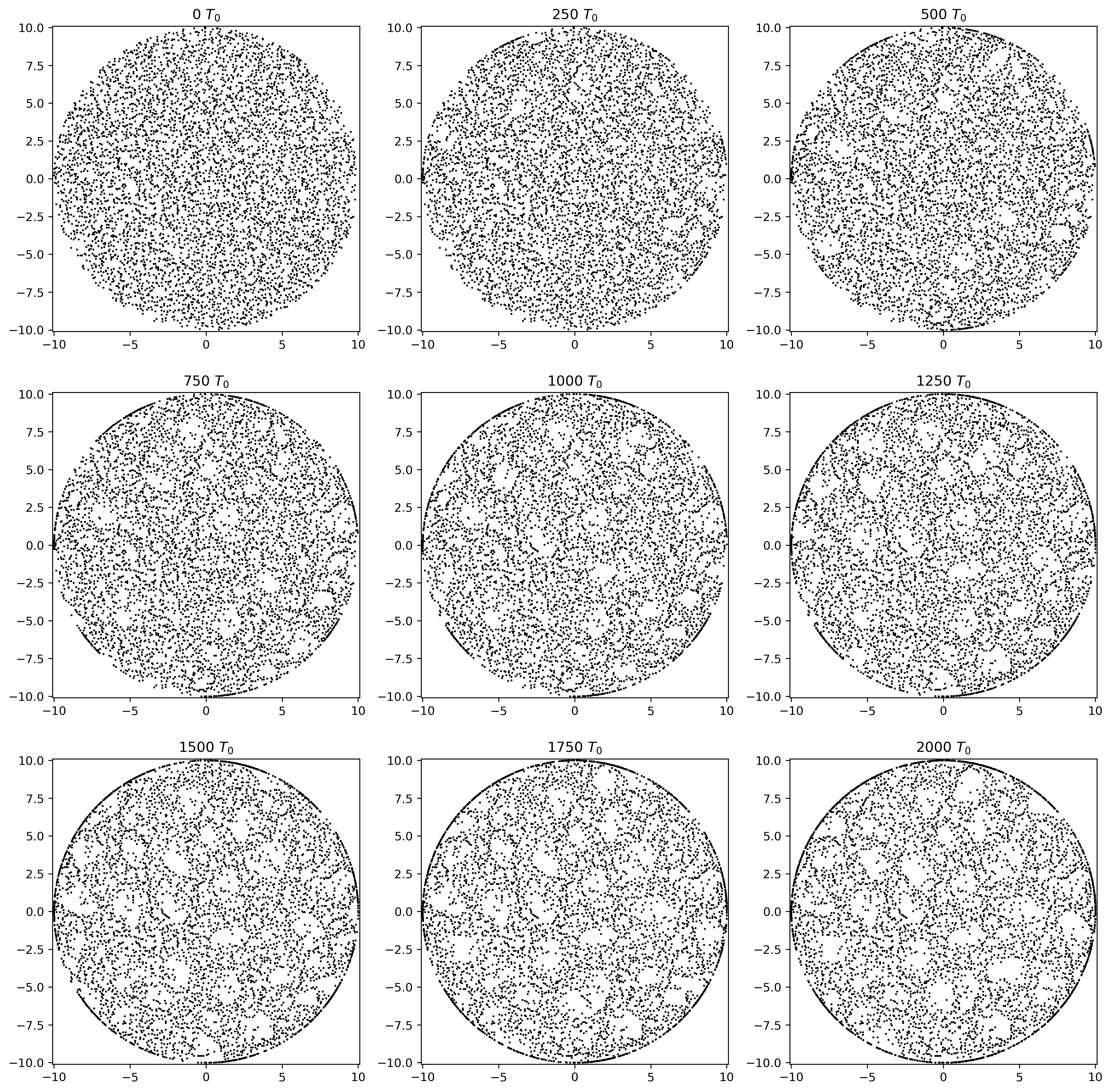}
  \caption{Evolution of the vortex array in a star from 0 $T_0$ to 2000 $T_0$, a duration comparable to the spin-down time. The panels are arranged chronologically, left to right, and top to bottom, beginning from the top left. The initially uniform array develops a few vacancies which grow in size while allowing for the nucleation and growth of other vacancies. Towards the last panel, voids of a similar size become apparent.}
  \label{fig:evolution_array}
\end{figure}

When we examine the evolution of an initially uniform vortex array in a spinning down neutron star (with pinning sites distributed uniformly as a square lattice and no assigned trap structure), we find that significant inhomogeneities in the vortex distribution can develop and sustain for long periods.  Fig.~\ref{fig:evolution_array} illustrates this phenomenon for a star initially populated with 5000 vortices and having $10^5$ pinning sites that can sustain a lag of up to 6.6 percent of the star's initial rotation rate. Pockets (vortex-free regions) are seen to develop spontaneously during glitches and continue to remain as vortex-free regions on timescales comparable to the spin-down timescale. A star with pinning sites that can support a greater critical unpinning speed (given by $v_{cr} = V_{0} \xi e^{-1/2}$) has fewer but larger pockets.

A vortex array that is initially uniform has largely been expected to remain so throughout the spin-down. However, we see that vortex-free regions having a well-defined length scale emerge spontaneously.\footnote{This is unlike the case of the vortex traps discussed in Section 3 where a length scale is deliberately introduced to mimic crustal rearrangement.} A detailed discussion of this will be presented in a forthcoming article.

\section{Conclusions}

We have investigated a model star that includes both traps and quakes that trigger the release of vortices from traps. We find that the trap structure can leave its signature on the rise profile of a glitch as a series of staggers. Additionally, we observe that all the glitches in our simulations have a fast rise compared to the spin-down timescale. In reality, the time resolution of the glitch rise is constrained by the duration of observation required to extract an accurate rotation frequency from a series of pulses having fluctuating shape. This introduces difficulties in detecting the staggers suggested by our simulations. There are however a few instances of glitches observed to have a seemingly staggered rise (\cite{lyne1992, wong2001, shaw2018}).

The distribution of glitch sizes is an equally viable object of study, being robust to low resolution data. The trap network does not seem to leave an immediate statistical imprint compared to the case without traps. However, the triggering of traps does. In some cases, this results in the distribution being bimodal, where the peak at the lower end occurs primarily due to glitches driven by spin-down alone. The location of the peaks could help us differentiate and narrow down the possible set of mechanisms at work. It must be reiterated that this bimodality occurs in the observationally relevant tail of the simulated distribution and does not span several orders of magnitude, as is the case in the Vela and Crab pulsars, where the star has a preference for larger or smaller glitches, respectively.

To faithfully understand individual pulsars, it is clear that we must place the glitch-size distribution on the same footing as the oft considered rise and relaxation profiles. A bottleneck to this is the small number of recorded glitches for most pulsars, which does not allow us to establish a reliable distribution. However, works that address this shortcoming by accumulating glitches from all pulsars also indicate a bimodality in the glitch-size distribution (\cite{konar2014, fuentes2017, eya2019}). A grouping of pulsars based on their age will help us better assess the distribution and its evolution.

This study, including the spontaneous development of macroscopic pockets discussed in Section 5, also highlights the need for efficient N-body simulations of large superfluid vortex arrays, an issue we have tried addressing in our newly established setup. However, our assumptions here neither account for variations in the nature of pinning across the layers of the star nor consider its effect on the rectilinear vortex array. Such details, which have previously been explored by others (\cite{levin2023,sheng2026}), could play a decisive role in where a glitch manifests, how it propagates, and what observational signatures it would produce.

\section*{Acknowledgements}

SVA and DB thank Ashoka University and Axis Bank for continued support. EG is supported by the Doctor Foundation of Qingdao Binhai University (No. BJZA2025025). The simulations in this paper were run, at various stages, on three different machines at Ashoka University; Chanakya HPC, Aryabhatta (courtesy Prof. Suratna Das, access facilitated by Umang Kumar), and Tao (courtesy Prof. Garima Mishra, access facilitated by Souradeep Sengupta). The animations were prepared using the Manim package available for python (\cite{manim2024}).

\section*{Data availability}

The simulation data used in this article can be made available upon request by emailing the corresponding author. However, the codes developed by the authors have been made publicly available through the following github repository: \href{https://github.com/ananth-94/vortexsimulator/tree/9f5a9f8955d34901221c0d98e56ab679645d8a07/superspin}{Superspin vortex simulator}. They can be used by appropriately citing this article.

%%%%%%%%%%%%%%%%%%%% REFERENCES %%%%%%%%%%%%%%%%%%

% The best way to enter references is to use BibTeX:

\bibliographystyle{mnras}
\bibliography{references} % if your bibtex file is called example.bib

% Alternatively you could enter them by hand, like this:
% This method is tedious and prone to error if you have lots of references
%\begin{thebibliography}{99}
%\bibitem[\protect\citeauthoryear{Author}{2012}]{Author2012}
%Author A.~N., 2013, Journal of Improbable Astronomy, 1, 1
%\bibitem[\protect\citeauthoryear{Others}{2013}]{Others2013}
%Others S., 2012, Journal of Interesting Stuff, 17, 198
%\end{thebibliography}

%%%%%%%%%%%%%%%%% APPENDICES %%%%%%%%%%%%%%%%%%%%%
\appendix

\section{Parameters of the simulations}

\begin{table*}
  \caption{Summary of the parameters common to all the simulations involving traps described in this article.}
  \label{tab:parameters}
  \begin{tabular}{@{}lll@{}}
  \toprule
  \textbf{Standard parameters} & \textbf{} &  \\ \midrule
  \begin{tabular}[c]{@{}l@{}}Radius of  the star\\ $R = 10$\\ \\ Number of  vortices \\ $N_v = 5\times10^{3}$\\ \\ Dissipation strength\\ $\phi = 0.1$\\ \\ Ratio of superfluid/crust\\ moments of inertia\\ $\mathrm{I_{rel}} = 1$\\ ~\end{tabular} & \begin{tabular}[c]{@{}l@{}}Number of pinning sites\\ $N_{\mathrm{pin}} = 10^{5}$\\ \\ Pinning strength\\ $V_0 = 10^{4}$\\ \\ Influence of pinning site\\ $\xi = 2.8\times10^{-3}~R$\\ \\ Runtime of simulation\\ $T_{\mathrm{run}} = 2000~ T_0$\\ ~\\ ~\end{tabular} & \begin{tabular}[c]{@{}l@{}}Initial rotation rate\\ $\Omega_0 = 40$\\ \\ Initial rotation period\\ $T_0 = 2\pi/\Omega_0 = 0.157\times10^{-3}$\\ \\ Spin-down deceleration\\ $N_{\rm{ext}}/I_{c} = -2.5 \times 10^{-4} ~\Omega_0/T_0$\\ \\ Default integration timestep\\ $dt = 0.005~T_0$\\ ~\\ ~\end{tabular} \\ \bottomrule
  \end{tabular}
\end{table*}

In this work, we simulate a star of radius 10 units, containing $5\times10^3$ quantized vortices, and $10^5$ pinning sites. We run our simulations in a dimensionless coordinate system where the length unit is one, the quantized circulation is $2\pi\kappa$, where $\kappa = 1$, and the crustal moment of inertia $I_{c} = 1$. All other quantities are defined through this fundamental set. Table~\ref{tab:parameters} summarizes the parameters relevant to the simulations described in Section 3 and Section 4. Below are a few comments on selected parameters and their values.

\subsection{Pinning}

Each pinning site is described by a Gaussian potential. That is, a vortex near a pinning site picks up a velocity given by:
\begin{equation}
    \mathbf{v}_{\text {pin }}=V_0 \exp \left[-\frac{\left(\mathbf{x}-\mathbf{x}_{\rm{pin}}\right)^2}{2 \xi^2}\right]\left(\mathbf{x}-\mathbf{x}_{\rm{pin}}\right) \times \mathbf{e}_z,
\end{equation}
where $\mathbf{x}$ and $\mathbf{x}_{\rm{pin}}$ are the position vectors of the vortex and the pinning site, respectively. The effect of a site is to rotate a nearby vortex in a clockwise fashion, opposite to the counter-clockwise effect of one vortex on another. $V_0$ is referred to as the pinning strength and $\xi$ is the characteristic pinning radius beyond which the effect of the site decays rapidly. We notice that if we choose $V_0 =10000$ and $\xi = 0.1~a$, where $a$ is the distance between neighbouring pinning sites, most of the vortices in a star with a uniform distribution of pinning sites get pinned by the end of the relaxation phase, and exhibit glitches during the spin-down phase. 

\subsection{Pinning in traps}
A trap structure is created by endowing a fixed non-zero pinning strength, $V_{0} = 10000$, to all pinning sites that fall within designated trap regions. Pinning sites located outside are assigned a pinning strength of zero. If all the vortices that were earlier distributed across the star were now initialized within the traps, it is seen that several of them are forced out of traps and into the vortex-free regions, by the end of the relaxation phase. The pinning is simply insufficient to arrest the increased density of vortices within the traps. To rectify this, we take the range of each pinning site to be half the distance between neighbouring sites, $\xi = 0.5~a$. While this is not reflective of an individual site in a neutron star, the fact that the number of such sites is tremendously large in the star essentially places the vortex in a region of high capture probability.

% Don't change these lines
\bsp	% typesetting comment
\label{lastpage}
\end{document}